\documentclass[11pt]{article}
\newcommand{\be}{\begin{eqnarray}}\newcommand{\beq}{\begin{equation}}
\newcommand{\ee}{\end{eqnarray}}\newcommand{\eeq}{\end{equation}}
\newcommand{\ep}{\varepsilon}\newcommand{\De}{\Delta}
\newcommand{\la}{\lambda}
\input epsf
\input amssym.def
\input amssym.tex
\usepackage{graphicx} 
\title{Effect of the surface-stimulated mode on the 
kinetics of homogeneous crystal nucleation in droplets} 
\author{ Y. S. Djikaev,$^a$\thanks{E-mail: idjikaev@eng.buffalo.edu}\hspace{0.19cm}\\ 
$^a$Department of Chemical and Biological Engineering, SUNY at Buffalo
\\ Buffalo, NY  14094 
\\.\\ 
}
\date{(Received\hfill }
\renewcommand{\baselinestretch}{2}
\begin{document}
\renewcommand{\baselinestretch}{1}
\maketitle
\renewcommand{\baselinestretch}{1}
{\bf Abstract.}
\renewcommand{\baselinestretch}{1} 
{\small

A kinetic theory of homogeneous crystal nucleation in unary droplets is presented taking
into account that a crystal nucleus can form not only in the volume-based mode (with all its
facets {\em within} the droplet) but also in the surface-stimulated one (with one of its
facets {\em at} the droplet surface). The previously developed thermodynamics of
surface-stimulated crystal nucleation rigorously showed that if at least one of the facets
of the crystal is only partially wettable by its melt, then it is thermodynamically more
favorable for the nucleus to form with that facet {\em at} the droplet surface rather than
{\em within} the droplet. So far, however, the kinetic aspects of this phenomenon had not
been studied at all. The theory proposed in the present paper advocates that even in the
surface-stimulated mode crystal nuclei initially emerge (as  sub-critical clusters) 
homogeneously in the sub-surface layer, not ``pseudo-heterogeneously"  at the surface. A
homogeneously emerged sub-critical crystal can become a surface-stimulated nucleus due to
density and structure fluctuations. This effect contributes to the total rate of crystal
nucleation (as the volume-based mode does). An explicit 
expression for the total per-particle rate of
crystal nucleation is derived. Numerical evaluations for water droplets
suggest that the surface-stimulated mode can significantly enhance the per-particle rate of
crystal nucleation in droplets as large as $10$ $\mu$m in radius. Possible experimental
verification of the proposed theory is discussed. 

}
\renewcommand{\baselinestretch}{2}
\newpage
\section{Introduction}
\renewcommand{\baselinestretch}{2}

\par The composition, size, and phases of aerosols and cloud particles determine their
radiative and chemical properties$^{1,2}$ thus determining the extent to and manner in which
the Earth climate is affected.  On the other hand, the composition, size, and phases of
atsmospheric particles are determined by the rate at  and mode in which these 
particles form and evolve.$^{2-4}$ Water constitutes an overwhelmingly dominant chemical species
participating in these atmospheric  processes hence a great importance attributed to studying
aqueous aerosols and cloud droplets as well as   their phase transformations. 

Although many phase transitions in aqueous aerosols  and cloud droplets occur via
heterogeneous nucleation on preexisting solid particles,$^{}$ in a number of important cases
atmospheric droplets appear to freeze homogeneously.$^{2,3,5}$ For example, the conversion of
supercooled water droplets into ice at temperatures below about -30$^{o}$C is known to occur
homogeneously, mainly because the concentrations of the observed ice particles in the clouds
often exceed the number densities of preexisting particles capable of nucleating ice.$^{6,7}$

Crystallization process in pure systems had long been assumed to initiate within the volume of
the supercooled phase.$^{8,9}$  Under that assumption, the rate of crystallization of a
droplet is proportional to its volume.$^{2,3,8,9}$  However, using  the classical nucleation
theory (CNT) based on the capillarity approximation,$^{10}$ we recently developed$^{11,12}$  a
thermodynamic theory that  prescribes the condition under which the surface of a droplet can
stimulate crystal nucleation therein so  that  the formation of a crystal nucleus with one of
its facets at the droplet surface is thermodynamically favored over its formation with all the
facets {\em within} the liquid phase.  This condition has the form of an inequality which,
when satisfied, predicts that  crystal nucleation in the droplet occurs mostly in a
``surface-stimulated" mode rather than in a ``volume-based" one.  For both  unary$^{11}$ and
multicomponent$^{12}$ droplets 
the inequality coincides with the 
condition for the partial wettability of at least one of the facets of a crystal nucleus by 
its own melt.$^{13}$ This effect was experimentally observed for 
several systems,$^{14,15}$ including water-ice$^{16}$ at temperatures at or 
below 0$^{o}$C. 

\par Although kinetic factors may play as important a role as thermodynamic ones in 
determining the mode of crystal nucleation, the partial wettability of a solid by its melt
may help to explain why, in molecular dynamics simulations$^{}$ of 
various kinds of supercooled liquid clusters$^{17,18}$ 
crystal nuclei appear preferentially  very close to the surface. 
As a result,  since smaller clusters have a higher surface-to-volume ratio, nucleation rates
in smaller clusters tend to be higher than in the bulk. Hence  it is experimentally easier to
observe the crystallization of aerosols,  having a large collective surface area, than those
having a large volume. The analysis$^{19,20}$  of laboratory data provided by various authors
also suggests  that the nucleation of both ice in supercooled water droplets and  nitric acid
hydrates in concentrated aqueous nitric acid droplets may initiate at the droplet surface
layer. Recent experiments$^{21,22}$  on the heterogeneous freezing of water droplets in both
immersion and contact modes  have also provided evidence that the rate of crystal nucleation in
the contact mode is much higher because the droplet surface may stimulate 
heterogeneous  crystal nucleation in the way similar to the enhancement of the homogeneous
process. 

It is well known that under otherwise identical thermodynamic conditions the free energy
barrier of heterogeneous nucleation is usually much lower than that of homogeneous 
nucleation,$^{13}$ so it might seem that  the idea of surface-stimulated crystal nucleation
does not significantly contribute to clarifying the underlying physics of the crystallization
phenomenon. In this regard, it should be noted that the surface-stimulated crystallization is
{\em not} a particular case of  heterogeneous nucleation. On the contrary,  it is a particular
case of homogeneous crystal nucleation hence its apparent thermodynamic similarities with
heterogeneous nucleation should be interpreted with due caution. Our analysis  in the present
paper will show that the kinetics of this process cannot be treated by using the formalism of
heterogeneous nucleation on foreign surfaces. 

\par The paper is structured as follows. In section 2 we briefly outline main
results$^{11,12}$  concerning the thermodynamics of surface-stimulated crystal nucleation
occurring homogeneously. For the sake of simplicity, in this work we consider only unary
systems, i.e., pure water droplets, but the generalization to  multicomponent droplets can be
carried out in the same manner as the unary theory in ref.11 was extended to multicomponent
systems in ref.12. A kinetic theory of such a process is presented in section 3. Numerical
predictions and possible experimental verification of the model are discussed in Section 4.
The results and conclusions are summarized in section 5. 

\section{Thermodynamics of surface-stimulated crystal nucleation}

To determine the conditions under which the surface of a droplet can thermodynamically 
stimulate its crystallization, it is  necessary to consider the formation of a crystal cluster
a) within a  liquid droplet and b) with one of the crystal facets at the liquid-vapor
interface (Figure 1). The criterion for the surface-stimulated crystallization is obtained by
comparing the reversible works of formation of a crystal nucleus (critical cluster) in these
two case. This was done in the framework of CNT  for both unary$^{11}$ and
multicomponent$^{12}$ droplets. The main results for unary droplets are outlined in this
section. 

\par Assuming the crystallization process to be isothermal and  neglecting the density
difference between liquid and solid phases, one can say that the total volume $V$, temperature
$T$, and  number of molecules $N$ in the system will be constant.$^{13,23}$  Then the 
reversible work of formation of a crystal cluster, $W$, can be found as  a change in the
Helmholtz free energy of the system upon its transition from the initial state (vapor+liquid)
into the final one (vapor+liquid+crystal).$^{11,12}$ 

\par Consider a crystal cluster (phase $s$) formed {\em within} a liquid droplet 
(phase $l$) which is surrounded by the vapor (phase $v$) (see 
Figure 1). The single superscripts $l,\;s$, and  $v$ will denote quantities in the
corresponding phases, whereas double 
superscripts will denote quantities at the corresponding interfaces. 
The reversible  work of formation of a crystal of
arbitrary shape with $n$ facets is 
\beq W=\nu[\mu^{s}(P^{s},T)-\mu^{l}(P^{l},T)]-
V^{s}(P^{s}-P^{l})+\sum_{i=1}^{\lambda}
\sigma^{ls}_{i}A^{ls}_{i}.\label{W1},\eeq
where 
$\nu$ and $\mu^{s}(P^{s},T)$ are the number of
molecules and chemical potential in the solid particle
formed within the liquid,
$P^{s}$ being the pressure within the
crystal and  $V$ its volume;  
$\mu^{l}(P^{l},T)$ is the chemical potential in the remaining liquid with pressure $P^{l}$,
$\sigma^{ls}_{i}$ and $A^{ls}_{i}$ are the surface tension and surface area of 
facet $i\;\;\;(i=1,...,n)$ of the crystal particle (anisotropic interfacial
energies are crucial in determining the
character of the nucleation process). 
By definition,$^{13}$ the surface tension of the solid is equal to the surface free energy
per unit area if the adsorption at
the solid-fluid interfaces is negligible, which usually is a reasonable assumption.
The pressure in the droplet is related to the pressure $P^{\beta}$ 
in the surrounding vapor via the Laplace equation $P^{l}=P^{v}+2\sigma^{lv}/R$, with 
$\sigma^{lv}$ being the droplet surface tension and $R$ 
the droplet radius (considered constant during crystallization).

\par Assuming that in the temperature
range between $T$ and the bulk melting temperature $T_{0}$  
the enthalpy of fusion does not change 
significantly, one can rewrite equation (1) as  
\beq W=-\nu\De h\ln\Theta+\sum_{i=1}^{n} \sigma^{ls}_{i}A^{ls}_{i},\eeq
where $\De h<0$ is the enthalpy of fusion (see, e.g., ref. 23), and $\Theta=T/T_{0}$. 
Note that in eq.(1) the mechanical effects
within the crystal (e.g., stresses) are considered to reduce to an isotropic pressure
$P^{s}$, so that$^{8,13}$ 
\beq 
P^{s}-P^{l}=
\frac{2\sigma^{ls}_{i}}{h_{i}}\;\;\;(i=1,\ldots,n),
\label{equcond-1}\eeq
where $h_{i}$ is the distance from facet $i$ to a point $O$ within
the crystal such that (see Figure 2) 
\beq \frac{\sigma^{ls}_{1}}{h_{1}}=\frac{\sigma^{ls}_{2}}{h_{2}}=\ldots
=\frac{\sigma^{ls}_{n}}{h_{n}}.\eeq
These equalities represent the
necessary and sufficient conditions for the equilibrium shape of the
crystal. This is known as the Wulff form and the equalities
themselves are Wulff's relations (see,
e.g., refs.8 and 13).
\par Equation (3) applied to the crystal is the equivalent of Laplace's equation applied to
liquid. Thus, just as for a droplet, one can expect to find a high pressure within a small
crystal.  Using eqs.(3) and (4), one can show that, for a crystal surrounded by the liquid
phase (melt),$^{11,12}$ 
\beq V^{s}(P^{s}-P^{l})=\frac2{3}\sum_{i=1}^{n} \sigma^{ls}_{i}A^{ls}_{i}.\eeq

\par By definition, the critical crystal (i.e., nucleus) is 
in equilibrium with the surrounding melt. For such a crystal
the first term in eq.(1) vanishes. Therefore, by virtue of
eq.(5), $W_{*}$, the reversible work of formation of the
nucleus, can be written as 
\beq W_{*}=\frac1{2}V^{s}_{*}(P^{s}_{*}-P^{l})=\frac1{3}
\sum_{i=1}^{n} \sigma^{ls}_{i}A^{ls}_{i*},\eeq
where the subscript ``*'' indicates quantities corresponding to the
critical crystal (nucleus). 

\par Now consider the case where a crystal cluster forms with 
one of its facets (say, facet $\lambda$) at a droplet surface.  Assuming that $
A^{sv}_{\la}/\pi R^{2} \ll 1$, the  deformation of the droplet can be neglected$^{11}$
and  the reversible work of formation of a crystal 
particle with its facet $\la$ at the droplet surface is 
\beq \widetilde{W}=\nu[\mu^{s}(P^{s},T)-\mu^{l}(P^{l},T)]-\widetilde{V}(P^{s}-P^{l})
+{\sum_{i=1}^{n}}^{\la}\sigma^{ls}_{i}A^{ls}_{i}
+\sigma_{\la}^{vs}A_{\la}^{vs}
-\sigma^{vl}A_{\la}^{vs},\eeq
where $\widetilde{V}$ is the volume of the crystal and where ${\sum }^{\la}$ 
indicates that the term with $i=\la$ is excluded from the sum. 

\par Wulff's relations in eq.(4), which determine the equilibrium shape of a
crystal, can be regarded as a series of equilibrium conditions on the
crystal ``edges'' formed by adjacent facets. For example, on the edge
between facets $i$ and $i+1$ the equilibrium condition is
$
\frac{\sigma^{ls}_{i}}{h_{i}}=
\frac{\sigma^{ls}_{i+1}}{h_{i+1}}.$
In the case where one of the facets (facet $\la$)
is the crystal-vapor interface while 
all the others lie within the liquid phase (see Figure 2), the
equilibrium conditions on the edges formed by this facet with the
adjacent ones (hereafter marked by a subscript $j$) are given by
\beq \frac{\sigma^{ls}_{j}}{h_{j}}=
\frac{\sigma^{sv}_{\la}-\sigma^{lv}}{\widetilde{h}_{\la}}.\eeq
Note that the height of the $\la$-th pyramid
(constructed with base on facet $\la$ and with apex at point $O$
of the Wulff crystal) will differ from that with all of
the facets in the liquid. Thus, the shape of the crystal
will differ from that in which all facets are in contact with the
liquid. For this case, Wulff's relations take the form
\beq \frac{\sigma^{ls}_{1}}{h_{1}}=\ldots=\frac{\sigma^{ls}_{\la-1}}{h_{\la-1}}=
=\frac{\sigma^{ls}_{\lambda}-\sigma^{lv}}{\widetilde{h}_{\lambda}}=
\frac{\sigma^{ls}_{\la-1}}{h_{\la-1}}=\ldots=\frac{\sigma^{ls}_{n}}{h_{n}},\eeq
and eq.(3) becomes
\beq P^{s}-P^{l}=
\frac{2\sigma^{ls}_{i}}{h_{i}}\;\;\;(i=1,\ldots\la -1,\la+1,\ldots,n),\;\;\;
P^{s}-P^{l}=
\frac{2(\sigma^{sv}_{\la}-\sigma^{lv})}
{\widetilde{h}_{\la}}.\eeq
Making use of equations (9) and
(10), one can represent eq.(7) as
\beq \widetilde{W}=-\nu\De h\ln\Theta
+{\sum_{i=1}^{n}}^{\la}\sigma^{ls}_{i}A^{ls}_{i}+
\sigma_{\la}^{vs}A^{vs}_{\la}
-\sigma^{vl}A_{\la}^{vs}.\eeq
\par For a crystal with one of
its facets a solid-vapor interface, and the others interfaced 
with the liquid, one can show that
\beq \widetilde{V}^{}(P^{s}-P^{l})=\frac2{3}\left({\sum_{i=1}^{n}}^{\la}
\sigma^{ls}_{i}A^{ls}_{i}+
\sigma_{\la}^{vs}A^{vs}_{\la}
-\sigma^{vl}A^{vs}_{\la}\right).\eeq
The reversible work $\widetilde{W}_{*}$ of formation
of a {\em critical} crystal can be thus represented as 
\beq \widetilde{W}_{*}=\frac1{2}\widetilde{V}^{}_{*}(P^{s}_{*}-P^{l})=
\frac1{3}\left({\sum_{i=1}^{n}}^{\la}
\sigma^{ls}_{i}A^{ls}_{i}+
\sigma_{\la}^{vs}A^{vs}_{\la}
-\sigma^{vl}A^{vs}_{\la}\right).\eeq

\par The similarity of equations (13) and (6) allows one to meaningfully compare them.
One can show$^{11,12}$ 
that the difference $P^{s}_{*}-P^{l}$ for the
nucleus is determined exclusively by the degree of 
supercooling of the liquid, so that in both eq.(6) and eq.(13)  
\beq (P^{s}_{*}-P^{l})_{}=
\frac{\De h}{v}\ln\Theta,\eeq
where $v$ is the volume per molecule in a solid phase. 
Using eqs.(3) and (10), one obtains 
\beq \widetilde{h}_{\la}=\frac{\sigma^{vs}_{\la}-\sigma^{lv}}
{\sigma^{ls}_{\la}}h_{\la}. \eeq
On the other hand,  $h_{i}=\widetilde{h}_{i}$ for $i=1,\ldots,\la -1,\la +1,\ldots, n$,
by virtue of eqs.(3), (10), and  (14). Therefore, the Wulff
shape of the surface-stimulated crystal is
obtained by simply changing the height of the $\la$-th pyramid of the volume-based Wulff
crystal. It is thus clear that if $\sigma^{sv}_{\la}-\sigma^{lv}<
\sigma^{ls}_{\la}$, then $\widetilde{h}_{\la}<h_{\la}$ and hence
$\widetilde{V}_{*}<V_{*}$.
Since
\beq \frac{\widetilde{W}_{*}}{W_{*}}=\frac{\widetilde{V}_{*}}{V_{*}}\eeq 
(according to eqs.(6),(13) and (14)), we can conclude that if 
\beq \sigma^{sv}_{\la}-\sigma^{lv}<
\sigma^{ls}_{\la},\eeq
then $\widetilde{W}_{*}<W_{*}$. In other words, if condition (17) is
fulfilled, it 
is thermodynamically more favorable for the crystal nucleus to form with facet $\la$ {\em at}
the droplet surface rather than {\em within} the droplet.

\par Inequality (17) coincides with the condition of partial wettability of the
$\la$-th facet of the crystal by its own liquid phase$^{13}$ (note that it has exactly the
same form in multicomponent droplets$^{12}$).  This result is physically reasonable,
because,$^{24}$ if the condition of partial wettability holds, the free energy per unit
area required to form a {\em direct} interface between  bulk vapor and solid (as in case of
surface-stimulated crystallization)  is less than the free energy required to  form a uniform
{\em intruding} layer of liquid phase, which involves creation of two interfaces
``solid-liquid  and ``liquid-vapor".  This intuitive argument alone, however, is {\em not
sufficient} to claim that the droplet surface stimulates crystal nucleation,  because (as
shown in ref.11 and 12) the volume-located  and surface-faced nuclei {\em differ} from each
other in shape and size (and, possibly, composition). It is necessary to compare the total
surface contributions to the free energy of nucleus formation rather than the free energies
per unit area of one particular facet of the nucleus, i.e., it is  necessary to compare the
quantity 
$\frac1{3}\left({\sum_{i=1}^{n}}^{\la}\sigma^{ls}_{i}A^{ls}_{i}+
\sigma_{\la}^{sv}A^{sv}_{\la} -\sigma^{lv}A^{sv}_{\la}\right)$ to 
$\frac1{3}\sum_{i=1}^{n} \sigma^{ls}_{i}A^{ls}_{i}$ 
rather than quantity  
$\sigma_{\la}^{sv}A_{\la}^{sv}-\sigma_{\la}^{lv}A_{\la}^{sv}$ to   
$\sigma^{ls}_{\la}A_{\la}^{ls}$. It is a mere coincidence that the first comparisons
reduces to the second one. 

\section{Kinetics of surface-stimulated crystal nucleation}

\par Inequality (17) allows one to predict whether crystallization in a 
supercooled droplet will or will not be thermodynamically stimulated by the surface.  To
apply this in practice, however, one needs accurate and detailed information about 
the surface tension of the liquid-vapor interface and the surface tensions of crystal
facets both in the liquid and in the vapor. Data on $\sigma^{lv}$ are available for most liquids
of interest or can be easily obtained. The availability of  data on $\sigma^{ls}$ and
$\sigma^{sv}$ is more problematic. Data on $\sigma^{ls}$ are often obtained by matching
experimental crystal nucleation rates with the predictions of CNT, 
treating the surface tension of the crystal nucleus as an adjustable
parameter.$^{9,25}$  However, such data are not suitable for 
using in eq.(17) for several reasons (see ref.11,12 for more details), one of which is the 
unsatisfactory state of the kinetic theory of crystal nucleation in light of the results
outlined in the previous section. 

\par Indeed, the classical expression for the rate of crystal nucleation conventionally used
in atmospheric models as well as for treating experimental data,  is derived by assuming that
crystal nuclei form {\em within}  the liquid.$^{8,9,2,3}$ However, under conditions of 
partial wettability of at least one crystal facet by its melt, the formation of a crystal
nucleus with that facet at the droplet surface is thermodynamically favored over its formation
with all the facets  {\em within} the droplet. This effect can become important when the
crystallizing liquid  is in a dispersed state, which is the case with the freezing of
atmospheric droplets$^{2,3}$ and many experiments.$^{26}$ 

Assuming a monodisperse (or sharp enough Gaussian) distribution  of liquid droplets, the
average crystallization time of the ensemble equals  that of a single droplet (for simplicity,
hereafter we will discuss only the monodisperse distribution, although results will be also
applicable to a narrow enough Gaussian-like  distribution).  Let us denote that time by $t_1$.
Typical sizes  of  atmospherically relevant droplets allow one to assume that the formation of
a single crystal nucleus in a droplet immediately leads to the crystallization of the latter,
i.e., the time of growth of a crystal nucleus to the size of the whole droplet is negligible 
compared to the time necessary for the first nucleation event in the droplet to occur (in
experiments this can be achieved by using special techniques$^{26}$). Consequently, 
\beq t_1=1/I, \eeq
where $I$ is the per-particle (pp) nucleation rate, i.e.,  the total number of crystal nuclei
appearing in the whole volume  of the liquid droplet per unit time. Until recently, atmospheric
models had considered  homogeneous crystal nucleation in droplets to be 
exclusively volume-based, with  
\beq I=I^{vb}\equiv J_vV_1, \eeq
where  $V_1$ is the volume of a single droplet and 
$J_v$ is the the rate of volume-base crystal nucleation given, e.g., by$^{9,2,3}$  
\beq J_v=\frac{kT}{h}\rho_l\mbox{e}^{-W_{*}/kT},\eeq 
with $k$ and $h$ being the Boltzmann and Planck constants and $\rho_l$ the number density of
molecules in the liquid phase. However, since the surface-to-volume ratio of a droplet increases with
decreasing droplet size,  eq.(19) may become inadequate for small enough
droplets in which the surface-stimulated crystal nucleation can compete with (or even dominate) 
the volume-based process. To take this possibility into account, it was recently$^{19,20}$ 
suggested
that instead of eq.(19) the pp-rate of crystal nucleation should consist of two contributions, 
\beq I=J_sS_1+J_vV_1,\eeq 
where  $S_1$ is the area of the droplet surface and $J_s$ is the number  of crystal nuclei forming
per unit time on unit surface area of the droplet (i.e., in a surface stimulated mode). 
The rate $J_s$ was
conjectured$^{19,20}$ to have the following form (reminiscent of that for the rate of crystal nucleation on
heterogeneous surfaces$^{2,3}$) 
\beq J_s=\frac{kT}{h}\rho^s_l\mbox{e}^{-\widetilde{W}_{*}/kT}, \eeq
with $\rho^s_l$ being the number of (liquid phase) molecules per unit area of the droplet
surface. 

\par As clear, using eq.(22) for $J_s$ on the RHS of eq.(21) amounts to implicitly assuming
that in the surface-stimulated mode the crystal nucleus forms in a heterogeneous fashion on a
molecule $m_s$ located in the droplet surface monolayer. As a first step, the first nearest
neighbors (both in the bulk and in the droplet surface layer)of molecule $m_s$  would acquire
a stable crystalline configuration (this would be due to fluctuations and hence might be
temporary). At the second step, the second nearest neighbors (including those in the droplet
surface layer) of molecule $m_s$ would acquire a stable crystalline order (this would be again
due to fluctuations and hence temporary). These steps would continue until the crystalline
cluster formed around $m_s$ attained a critical size (i.e., became a crystal nucleus with one
of its facets at the droplet-vapor interface) which would be followed by a quick
crystallization of the whole droplet. 
The number of nuclei forming per unit time per unit area
of the droplet surface (i.e., the surface nucleation rate $J_s$) would be then given by
eq.(22).  

\par However, there is a weak point in the above reasoning. Indeed, heterogeneous mechanism of 
nucleation implies that the {\em initial} stage of the formation of a new phase fragment
around a heterogeneous center is thermodynamically favorable, i.e., it is accompanied by a
decrease in the appropriate free energy of the system. This was clearly demonstrated for
heterogeneous condensation on  ions$^{27,28}$ as well as on insoluble,$^{29,30}$
mixed,$^{31}$ and various soluble$^{32}$ macroscopic particles.  For all these
phenomena a specific physical effect, causing the initial decrease in the free energy of the
system upon the formation of a new phase particle around a heterogeneous center, can be
unambiguously identified. This is not the case with surface-stimulated crystal nucleation in
droplets. The formation of a crystal nucleus with one of its facets at the droplet surface 
cannot {\em start} preferentially {\em at} the surface, because the latter does not have any
sites which would make the ordering of the surrounding {\em surface located}  molecules 
thermodynamically more favorable than the ordering of interior molecules. On the contrary, the
surface layer of a crystalline structure remains disordered far below the freezing/melting
temperature. This phenomenon is   often referred to as premelting$^{33}$ and was observed
both experimentally$^{34-36,15}$ and by simulations$^{37-40}$. 

According to the empirical criterion proposed by Lindemann,$^{41}$ 
melting in the bulk  can occur when the root mean
amplitude of thermal vibrations of an atom exceeds a certain threshold value 
of more or about $10\%$ of the distance to the nearest neighbor in the crystalline structure.
Developing this criterion, 
Tammann suggested$^{33}$ that the outermost  layer of the
crystal should become disordered far below the bulk melting point due to the higher freedom 
of motion for surface-located molecules which have a
reduced number of neighbors and hence have a higher vibrational 
amplitude compared to those in the bulk.  
Thus, one can expect the Lindemann criterion for surface atoms to be satisfied at a 
temperature lower than that for atoms located within the crystalline structure. 
The surface melting (often referred to as premelting) involves the formation of a thin 
disordered layer at a temperature significantly below the melting one. 

Experimentally the premelting phenomenon was apparently first detected by Lyon and
Somorjai$^{34}$ who studied the structures of clean (111),(110), and (100) crystal faces of
platinum as a function of temperature by means of low-energy electron diffraction and observed
the formation of disordered  surface structures  at temperatures far below the melting
temperature $T_m=2043$ K.  Direct experiments on the surface-initiated melting were also
carried out by  Frenken {\em et al.}$^{35}$ using Rutherford backscattering in conjunction
with  ion-shadowing and blocking. That experiment revealed a reversible order-disorder
transition on the (110) surface of a lead crystal well below its melting point $T_m=600.7$ K. 
Since  then, other techniques have been employed such as calorimetry, electron, neutron, and
X-ray diffraction, microscopy, ellipsometry, and helium scattering. Although most experiments
were  carried out under equilibrium conditions, melting tended to be initiated at the surface
even when  the crystalline solid was heated very quickly so that equilibrium conditions were
not  established.$^{36}$  Lately, molecular dynamics simulations have been also 
widely used to study 
premelting (see, e.g., ref.37 and recent simulations of premelting in  $AgBr$ (ref.38), in
$Cr_2O_3$ (ref.39), and the premelting of a clean $Al$(110) surface$^{40}$). 


Of utmost atmospheric relevance,   there has been accumulated undeniable evidence$^{16,42-44}$ for
the premelting of ice (first apparently discussed by Faraday$^{45}$). Relatively recently Wei
{\em et al.}$^{46}$ experimentally observed that the premelting of the (0001) face of hexagonal
ice occurs at the temperature of about $200$ K, i.e., much below the lowest temperature
reported for homogeneous freezing of atmospheric droplets.  Thus, although the droplet surface
can (under condition (17)) stimulate crystal nucleation, a crystal nucleus with one
facet as a droplet-vapor interface most likely {\em begins} its formation (as a subcritical
crystal)  homogeneously in a spherical layer adjacent to the droplet surface (``sub-surface
layer").  When this crystal becomes large enough (due to fluctuational growth usual for the
nucleation stage), one of its facets hits the droplet surface and at this moment or shortly
thereafter it becomes a nucleus owing to a drastic change in its thermodynamic state. 

Let us consider the ``sub-surface" layer of the droplet hereafter  referred to as an SSN layer (SSN
can stand for both ``sub-surface nucleation" and ``surface-stimulated nucleation"). Its thickness will
be denoted by $\eta$ (more detailed discussion of $\eta$  is given in the following subsection). By
definition (albeit somewhat loose for now), any  crystalline cluster that starts its evolution with
its center in the SSN layer has a potential to become a nucleus (by means of structural and density
fluctuations) once one of its facets that  satisfies the condition of partial wettability, eq.(17),
meets the droplet surface.  Clearly, to become a surface-stimulated nucleus, the subcritical cluster
must evolve  in such a way that its facet $\la$ (satisfying condition (17)) is parallel to the droplet
surface at the time they meet. The orientation adjustment cannot be  mechanical because this would
require relatively long time scales, but may, or may not, occur by means of appropriate  spatial
distribution of density and structure fluctuations around the cluster. 

In the framework of the SSN layer model,  the pp-rate of crystal nucleation
is given by the sum 
\beq I=J_v^sV_1^s+J_v(V_1-V_1^s)\eeq  
(rather than by eqs.(19) or (21)), where  
$J_v^s$ is the number of crystal nuclei forming in a surface stimulated mode per unit time 
in unit volume of the SSN layer whereof the total volume is $V_1^s$. 
Equivalently, 
\beq I=I^{ss}+I^{vb}, \eeq 
where 
\beq I^{ss}=(J_v^s-J_v)V_1^s\eeq
and $I^{vb}$ (defined by eq.(19)) are the contributions to the total pp-rate of crystal nucleation arising from the surface-stimulated and volume-based modes, respectively. As clear from the above discussion,  
\beq J_v^s=\frac{kT}{h}\rho_l\mbox{e}^{-\widetilde{W}_{*}/kT}, \eeq
where the number density of molecules in the droplet is assumed to be uniform up to the
dividing  surface,$^{13}$ 
in consistency
with the capillarity approximation.$^{10}$ In this approximation, the pre-exponential
factors in eqs.(20) and (26) are the same, which reflects the {\em homogeneous} nature of the
nascency of a sub-critical cluster in both volume-based and surface-stimulated modes of
crystal nucleation. On the other hand, the nuclei in these two modes are different hence have
different free energies of formation, which results in different exponents: $-W_{*}/kT$ for the
volume-based process and $-\widetilde{W}_{*}/kT$ for the surface-stimulated mode 
(see section 2). 

By virtue of eqs.(19),(20), and (26), one can rewrite equation (25) as 
\beq I^{ss}=\alpha I^{vb},\eeq
where   
\beq \alpha\equiv\alpha(\ep)\equiv \alpha(\ep,\De W_{*}(\eta(ep)))=[1-(1-\ep)^3](-1+\mbox{e}^{-
\De W_{*}/kT}) \eeq 
with $\ep=\eta/R$ ($0\le\ep\le 1$). As clear from eq.(27), the ratio $\alpha=I^{ss}/I^{vb}$ 
characterizes the relative intensity of surface-stimulated and volume based modes of crystal
nucleation in the droplet. If the volume-based process predominates over the
surface-stimulated nucleation (i.e.,$I^{ss}\ll I^{vb}$), then $\alpha\ll 1$. If the
surface-stimulated mode prevails over the volume-based one (i.e., $I^{ss}\gg I^{vb}$), then
$\alpha\gg 1$. In the cross-over regime, when the nucleation mode factor is roughly in the range 
$0.33 \lesssim\alpha\lesssim 3$, the contributions from both modes to the
total pp-rate of crystal nucleation are comparable to each other. The exact cross-over point is
given by the equality $I^{ss}=I^{vb}$. 

According to eq.(28), under given external conditions (temperature, pressure, etc...) the
value of the ``nucleation mode factor" $\alpha$ is determined by $\ep=\eta/R$, i.e., by the
size of the freezing droplet, $R$.   The geometric factor  $1-(1-\ep)^3$ in $\alpha$
monotonically increases from $0$ to $1$ with increasing $\ep$.  However,
the nucleation mode factor itself, $\alpha$, may have a more complicated  dependence on $\ep$
because of the exponential $\mbox{e}^{-\De W_{*}/kT}$ in which $\De W_{*}$ is intrinsically
related to  $\ep$ via $\eta$. 

\subsection{Nucleation mode factor} 

According to the definition of the SSN layer, its thickness, denoted by $\eta$, is  determined
by the shape and orientation of the crystal nucleus and the physical characteristics of the
crystal nucleus   {\em and} droplet. They also determine the free energy of nucleus formation
in both surface-stimulated and volume-based modes,  $\widetilde{W}_{*}$  and $W_{*}$,
respectively (the droplet surface tension is involved only in $\widetilde{W}_{*}$  but not in
$W_{*}$). For a given droplet in a given thermodynamic state, $\eta$ and  $W_{*}$  are
completely independent of each other.  However, both $\eta$ and $\widetilde{W}_{*}$ are
determined by the shape {\em and} orientation of the crystal nucleus. Hence the dependence 
$\De W_{*} = \De W_{*}(\eta)$ which, according to eq.(28), is likely to be a key factor in
determining the $\ep$-dependence of $I^{ss}$ (because $\ep=\eta/R$) and, ultimately, the
value of the nucleation mode factor $\alpha$.

Consider first the case where crystal clusters (including the critical one, nucleus, which
forms as a result of fluctuational growth of an initially subcritical cluster) have $n$
facets each and assume that only one of these facets, say, facet $\lambda$, satisfies the
condition of partial wettability, eq.(17). Let us introduce the unit vectors ${\bf n}_d$ and
${\bf n}_{\la}$ as the external normal vectors to the droplet surface and facet $\la$,
respectively. The angle $\Theta$ between ${\bf n}_d$ and ${\bf n}_{\la}$
determines the mutual orientation of the droplet surface and facet $\la$. Clearly, $0\le\Theta\le \pi$
with $\Theta=0$ corresponding to the $\la$-facet being parallel to the droplet surface and
$\Theta=\pi$ their being ``antiparallel". 

If the only possible orientation of a crystal cluster were the one with $\Theta=0$, then the
$\la$-facet of any surface-stimulated nucleus would be a part of the droplet surface. Consequently, the
thickness $\eta$ of the SSN layer would be equal to $\widetilde{h}_{\la}$ (the height of the
$\la^{\mbox{\small th}}$ pyramid, with facet $\la$ its 
basis and point $O$ its apex, see Figure 2) and eq.(27) could be written as 
\beq I^{ss}=[1-(1-\ep_{\la})^3](-1+\mbox{e}^{-\De W_{*\la}/kT})I^{vb},\eeq 
where $\ep_{\la}=\widetilde{h}_{\la}/R$, $\De W_{*\la}\equiv \widetilde{W}_{*\la}- W_{*}$, and 
$\widetilde{W}_{*\la}$ is the free energy of formation of
a surface-stimulated nucleus with facet $\la$ being a part of the droplet surface.  

In reality, however, the orientations of crystal clusters are randomly distributed in the range
from $0\le\Theta\le 1$. Assuming this distribution to be uniform (because there are no obvious
reasons for the contrary), its probability density is $ p(\Theta)=\frac1{\pi}$ 
with the normalization $\int_0^{\pi}d\Theta\,p(\Theta)=1$. 

Let $p_{\Theta\ep}(\Theta)$ be the probability density that a crystal cluster has an orientation
$\Theta$ and, at this orientation, the surface-stimulated nucleation occurs in a layer of
(dimensionless) thickness $\ep$. As mentioned above, for a given droplet under given external
conditions  $\ep$ can be a  function of only $\Theta$. Moreover, the droplet surface can stimulate the
nucleation of crystal clusters only at one single orientation, $\Theta=0$. Thus, 
\beq p_{\Theta\ep}(\Theta)=p(\Theta)\delta(\Theta)=\frac1{\pi}\delta(\Theta).\eeq

The contribution $I^{ss}$ from the surface-stimulated mode to the total pp rate of crystal
nucleation, $I$, is now obtained by averaging   $I^{ss}(\ep(\Theta))$, given by eqs.(27) and 
(28), over all the possible 
orientations of crystal clusters, i.e., as 
\beq I^{ss}=\int_{0}^{\pi}d\Theta\,[1-(1-\ep(\Theta))^3](-1+\mbox{e}^{-
\De W_{*}(\Theta)/kT})\,I^{vb}\,p_{\Theta\ep}(\Theta).\eeq
Taking into account eq.(30) and the equality $\ep(0)=\ep_{\la}$, one obtains 
\beq I^{ss}=\frac1{\pi}[1-(1-\ep_{\la})^3](-1+\mbox{e}^{-\De W_{*\la}/kT})I^{vb},\eeq
As clear, the ability of a crystal cluster in the SSN layer to appear with facet $\la$
not only parallel to the droplet surface but with any other orientation,  with 
$\Theta$ uniformly distributed from $0$ to $\pi$, decreases $I^{ss}$ (and hence 
$\alpha$) 
by a factor of $1/\pi$ compared to a hypothetical situation when 
all crystal clusters would evolve with their $\la$ facets parallel to the droplet surface, i.e.,
with $\Theta=0$.

Equation (32) is obtained for the case where every cluster of the nascent crystalline
structure has only one facet (facet $\la$) satisfying the condition of partial wettability,
eq.(17). In a more general situations, every cluster can have $w$ facets   ($1\le w\le N$)
partially wettable by the melt. Let these facets be numbered $1$ through $w$.
Every one of these facets contributes to the surface-stimulated mode of the pp-rate of crystal
nucleation in the droplet. Since these contributions  $I_{\la}^{ss}\;\;\;(\la=1,...,w)$ are
independent of one another, each of them is determined by eq.(32), so that the total
``surface-stimulated" contribution $I^{ss}$ to the pp-rate of
crystal nucleation will be given by the sum $\sum_{\la=1}^{w}I_{\la}^{ss}$, i.e., 
\beq I^{ss}=\sum_{\la=1}^{w}\frac1{\pi}[1-(1-\ep_{\la})^3](-1+\mbox{e}^{-\De W_{*\la}/kT})
I^{vbn}.\eeq 

\par In a rough approximation, one can assume that 
\beq \widetilde{h}_{0}\equiv \widetilde{h}_{1}\approx \widetilde{h}_{2}\approx\ldots  
\approx \widetilde{h}_{w} \eeq 
and 
\beq \widetilde{W}_{*0}\equiv \widetilde{W}_{*1}\approx \widetilde{W}_{*2}\approx\ldots
\approx \widetilde{W}_{*w}.\eeq 
The former assumption is reasonable if, for instance,  all crystal nuclei have globular (not
elongated) shape  with aspect ratios close to $1$, whereas the latter implies that the surface
tensions of facets $1,...,w$ do not differ much from one another. With such approximations,
equation (33) reduces to
\beq I^{ss}=\frac{w}{\pi} [1-(1-\ep_{0})^3](-1+\mbox{e}^{-\De W_{*0}/kT})I^{vb}, \eeq 
where $\ep_0=\widetilde{h}_0/R$ and $\De W_{*0}=\widetilde{W}_{*0}- W_{*}$.  This equation is
convenient for rough numerical evaluations of $I^{ss}$.  In a more complicated  case  where
assumption (34) is not acceptable, while approximate equalities in eq.(35) do hold, eq.(33)
acquires the form
\beq I^{ss}=\frac1{\pi}[w-\sum_{\la=1}^{w}(1-\ep_{\la})^3](-1+\mbox{e}^{-\De W_{*0}/kT})
I^{vb} ... . \eeq
where the factor in the square brackets is a relatively weak function of
$\ep_{\la}\;\;(\la=1,...,w)$ not exceeding $w$. Again, this form is more convenient than
eq.(33) to numerically evaluate the nucleation mode factor $\alpha$. 

\section{Numerical Evaluations and Experimental Perspective}
To illustrate the above theory by numerical evaluations, consider the freezing of water droplets
(surrounded by water vapor in air) at around $T=233$ K (i.e., about $-40^o$C). The 
homogeneous and isothermal character of freezing is assumed. 

As reported by Defay {\em et al.}$^{13}$, the rate of crystal nucleation in bulk supercooled 
water at this temperature is $7\times 10^{12} cm^{-3}s^{-1}$, with the nucleation
barrier height $W_*=45\;kT$, the average (over all crystal facets) 
surface  tension of liquid-solid (water-ice) interface
$\sigma^{ls}$  being about $20$ dyn/cm (Table 18.1 in ref.12). The surface  tensions of
liquid-vapor and solid-vapor (ice-water vapor) interfaces at $T=233$ K will be taken to be
$\sigma^{lv}=88$ dyn/cm and $\sigma^{sv}=103$ dyn/cm, respectively. 
All these values of $\sigma^{ls},\sigma^{lv},$ 
and $\sigma^{sv}$  are consistent with the data provided in ref.2. 

The wettability of a solid by a liquid (both in contact with a vapor) is determined by the
contact angle, defined as the angle between the tangents to the liquid-vapor and solid-liquid
interfaces at the three phase contact line. According to Young's relation,$^{13}$ 
which gives a connection between three
interfacial tensions and contact angle,  the above values of
$\sigma^{ls},\sigma^{lv},$ and $\sigma^{sv}$ would correspond to the contact angle $\beta\simeq
19.4^o$ (or $\cos{\beta}\simeq 0.943$). Therefore, at $T=233$ K at least some of (if not all) 
the facets of an ice crystal are only partially wettable by liquid water. This is consistent
with the experimental results of Elbaum {\em et al.}$^{16}$ who reported partial wettability of
the basal facets of hexagonal ice (Ih) at temperatures slightly below $0^o$C. In those
experiments,$^{16}$ when air was added to water vapor the  partial wetting of ice by water
transformed into complete wetting but only for some orientations.  Besides, the
wettability of solids by fluids usually decreases with decreasing temperature,$^{47,48}$ so 
one
can expect  that at temperatures far below 0$^{o}$C at least some facets of water crystals
remain only partially wettable even in the presence of air.  Moreover, according to Cahn's
theory,$^{47}$ perfect wetting of a solid by a liquid away from the critical point is not
generally observed, i.e., condition (17) should be fulfilled for any substance at 
sufficiently low temperatures.  That theory$^{47}$ can be also applied to the case where  the
solid is of the same chemical nature as the fluid phases  (it is only assumed that the surface
of the solid phase is sharp on an atomic scale and interactions between surface and fluid are
sufficiently short-range). Cahn's theory is inapplicable at temperatures close to  the fluid
critical point, but  temperatures involved in crystallization are usually far below that point.

For simplicity of numerical evaluations, let us assume that only the basal facet \{0001\} of the
hexagonal ice crystal is partially wettable by water at $T=233$ K. Denote the height of the
basal pyramid of the crystal cluster by $\widetilde{h}_b$ when the basal facet is at the droplet surface
and by $h_b$ when the entire crystal is immersed in the droplet.  According eq.(15),  
\beq \frac{\widetilde{h}_{b}}{h_{b}}=\frac{95-80}{20}=3/4.\eeq
Consequently, the ratio of the volumes of the surface-based and
volume-formed clusters is  
\beq \frac{\widetilde{V_*}}{V_*}\simeq \frac{\widetilde{h}_{b}+h_{b}}{2h_{b}}.\eeq 
By virtue of eq.(16), for the corresponding works of formation of crystal nuclei 
we obtain  
\beq \frac{\widetilde{W}_*}{W_*}\simeq 0.875,\eeq
i.e., $\widetilde{W}_*\simeq 39.3\,kT$. Thus, the decrease in the work of 
formation of the surface-stimulated  
crystal nucleus, as compared to that of the volume-based one, is $\De W_{*}\approx -5.7\; kT$.

Let us now evaluate the thickness of the SSN layer, $\eta$, which still depends on the size and 
habit of crystal nuclei even after averaging over $\Theta$ in eq.(31). 
According to eq.(6), the work of formation of a
volume-formed crystal nucleus can be rewritten in the form
\beq W_*=\frac1{3}\sigma_{ls}A_*, \eeq 
where $\sigma_{ls}$ is the average (over all the crystal facets) interfacial tension of the
nucleus and $A_*$ is its total surface area.$^{11,12}$ The total surface area of an Ih  
crystal (shaped as a right prism) is   $A=12h_ba+ 3\sqrt{3}a^2$, where $a$ is the side length of the basal facet (for 
a regular hexagon $a$ is also equal to its radius) and $h_b$ is the half-height of the prism. 

One can define the aspect ratio of a volume-formed Ih crystal as $\gamma=h_b/a$.  For the
formation of snow crystals from water vapor $\gamma$ is a complex non-monotonic function of both
temperature and water vapor saturation ratio.$^{49}$ Likewise, for
crystallization in liquid water $\gamma$ is a complex function of temperature and
pressure.$^{50}$  For this reason, let us consider two opposite cases, $\gamma=2$  and
$\gamma=0.5$, corresponding to column-like and plate-like crystals of $Ih$, respectively.  As
mentioned, for crystallization in pure water $W_*\approx 45\;kT$ and $\sigma_{ls}\approx 20$
dyn/cm at $T=233$ K. Therefore, eq.(41) leads to  $h_b\approx 6.9\times 10^{-8}$ cm 
for $\gamma=0.5$ and
$h_b\approx 17.3\times 10^{-8}$ cm for $\gamma=2$ which, according to eq.(38), 
correspond to the following values of the thickness of the SSN layer:
\beq \eta=\widetilde{h}_b\approx 5.2\times 10^{-8} \mbox{ cm}\;\;\;\;\;(\gamma=0.5),
\;\;\;\;\;\;\;\;\;\; \eta=\widetilde{h}_b\approx 12.9\times 10^{-8} \mbox{ cm}\;\;\;\;\;(\gamma=2).\eeq

The nucleation mode factor $\alpha$ can be estimated from eq.(36) with $w=2$. Its dependence on the
radius of the droplet is presented in Figure 3, where the dashed and solid curves correspond to
$\gamma=2$ and $\gamma=0.5$, respectively. As clear, the surface-stimulated mode can considerably  
enhance crystal nucleation in droplets with radii even exceeding $1$ $\mu$m. For example, for
droplets of radius $R=0.2$ $\mu$m the nucleation mode factor $\alpha=3.3$ for $\gamma=2$ and
$\alpha=1.4$ for $\gamma=0.5$, while for droplets of radius $R=2$ $\mu$m the nucleation mode factor
$\alpha=0.34$ for $\gamma=2$ and $\alpha=0.14$ for $\gamma=0.5$. These estimates suggest that
homogeneous crystal nucleation in water droplets with radii smaller than $R\approx 0.2$ $\mu$m occurs 
predominantly in the surface-stimulated mode, while the volume-based mode prevails in droplets with
radii greater than $R\approx 2$ $\mu$m. Both the surface-stimulated and volume-based modes apparently
provide contributions of the same order of magnitude to the total pp-rate of crystal nucleation in
droplets with radii in the range approximately from $0.2$ $\mu$m to $2$ $\mu$m. Similar evaluations
can be carried out for the case where all the facets of Ih crystals are only partially wettable. 
Assuming that for the Ih prism facets $\De W_*$ is approximately the same as for the basal ones, it is
clear from eq.(36) that in this case the  nucleation mode factor $\alpha$ may be greater than $1$ even
for droplets with $R>10$ $\mu$m (i.e., the surface-stimulated mode may dominate even for such
relatively large droplets).  

While the above numerical estimates are approximate because of insufficiently accurate data on the
interfacial tensions involved in the model,  laboratory  techniques currently available for studying
crystal nucleation in droplets  make it possible to carry out the experimental verification of the
above theory. Indeed, modern experimental methods$^{2,26}$ can provide data on the
dependence of the pp-rate of  crystal nucleation in droplets on their radius, i.e., $I^{\mbox{\small
}}$ as function of $R$. According to the above model, one  can expect that there must exist such a
constant $A$ that for $R\lesssim 1$ $\mu$m  the LMS fit of the experimental dependence 
$I^{\mbox{\small exp}}$ vs $R$ with the function $(1+A/R)BR^3$ ($B$ is another constant) is much more
accurate than with the function $BR^3$ and the inaccuracy of  the latter compared to the former should
be aggravating with decreasing $R$. 

\section{Concluding Remarks}

The thermodynamics of surface-stimulated crystal nucleation was previously developed for both unary$^{11}$ and multicomponent droplets$^{12}$ (for which  the theory is more complicated not only due 
to the presence of several components, but also due to the surface  
adsorption of all components as well as their dissociation into ions) in the framework of CNT. A criterion was found for when the surface of a droplet can 
stimulate crystal nucleation therein so  that  the formation of a crystal nucleus with one of
its facets at the droplet surface is thermodynamically favored (i.e., occurs in a surface stimulated mode) over its formation with all the
facets {\em within} the liquid phase (i.e., in a volume-based mode). For both unary and multicomponent droplets,  this criterion  coincides with the 
condition of partial wettability of at least one of the crystal facets by 
the melt. However, so far the kinetic aspects of this phenomenon had not been studied at all. 

In this paper we have presented a kinetic theory of homogeneous crystal nucleation in unary droplets  taking into account that a crystal nucleus can form not only in the volume-based mode 
but also in the surface-stimulated one. We have invoked experimental and simulations-based evidence  showing that surface-stimulated crystal nucleation is 
{\em not} a particular case of  heterogeneous nucleation. On the contrary,  it is a particular
case of homogeneous crystal nucleation hence its thermodynamic similarities with
heterogeneous nucleation can be misleading because the kinetics of this process cannot be treated by using the formalism of heterogeneous nucleation on foreign surfaces. 

Even in the surface-stimulated mode the crystal nucleus initially emerges (as 
a subcritical cluster)  homogeneously in the droplet sub-surface layer, not pseudo-heterogeneously 
at the droplet surface. A homogeneously emerged sub-critical crystal cluster can become a surface-stimulated nucleus when, after growing large enough owing to density and structure fluctuations, one of its facets meets the droplet surface and both are parallel to each other. This effect gives rise to an additional contribution to the total rate of crystal nucleation in a droplet (the conventional contribution arises from the volume-based crystal nucleation). We have derived an expression for the total per-particle rate of crystal nucleation in the droplet in the framework of CNT. The theory has been presented  only for unary 
droplets, but its generalization to  multicomponent droplets is possible although not straightforward.

As a numerical illustration of the proposed theory, we have considered crystal nucleation in water
droplets at $T=233$ K. Our results suggest that that the surface-stimulated mode can markedly enhance
the per-particle rate of crystal nucleation in water droplets as large as $10$ $\mu$m in radius. We
have also roughly outlined a simple way to carry out the experimental verification of the proposed
theory. 

However complex a theory of homogeneous crystal nucleation in droplets may be,  the presence of foreign 
particles,  serving as nucleating centers, makes the crystal nucleation phenomenon (and hence its
theory) even more involved. Numerous aspects of heterogeneous crystal nucleation still remain obscure. 
For example, it has been observed that the same nucleating center initiates the crystallization of a
supercooled droplet at a higher temperature in the contact mode (with the foreign particle just {\em
touching} the droplet surface)  than in the immersion mode (particle {\em immersed} in the
droplet).[2,5,21,22] Underlying physical reasons for this enhancement have remained
unclear, but as little as might be known about  the phenomenon of surface-stimulated (homogeneous)
crystal nucleation, it strongly suggests  that  the droplet surface can enhance heterogeneous
nucleation in a  way similar to the enhancement of the homogeneous process. The thermodynamics and
kinetics of heterogeneous crystal nucleation in droplets (in both contact and immersion modes) is  
the subject of our current research. 

\subsubsection*{}
{\em Acknowledgment} - {\small YSD thanks Raymond Shaw for helpful discussions.
}
\section*{References}
\begin{list}{}{\labelwidth 0cm \itemindent-\leftmargin}
\item $(1)$ {\it IPCC, Climate Change 2001: The scientific bases}. 
Inter government Panel on Climate Change; 
Cambridge University Press, Cambridge UK, 2001.
\item $(2)$ Pruppacher, H. R.; Klett., J. D. {\it Microphysics of clouds 
and precipitation}. (Kluwer Academic Publishers, Norwell, 1997).
\item $(3)$ N.H.Fletcher, {\it The physics of rainclouds}. (University Press, Cambridge, 1962).
\item $(4)$ Cox, S. K. {\it J. Atmos. Sci.} {\bf 1971}, {\it 28}, 1513.
\item $^{5}$W.Cantrell and A.Heymsfield, BAMS, {\bf 86}, 795 (2005).
\item $(6)$ Jensen, E. J.; Toon, O. B.; Tabazadeh, A.; Sachse, G. W.; Andersen,
B. E.; Chan, K. R.; Twohy, C. W.; Gandrud, B.; Aulenbach, S. M.;
Heymsfield, A.; Hallett, J.; Gary, B. {\it Geophys. Res. Lett.} {\bf 
1998}, {\it 25}, 1363.
\item $(7)$ Heymsfield, A. J.; Miloshevich, L. M. {\it J. Atmos. Sci.} 
{\bf 1993}, {\it 50}, 2335.
\item $^{8}$ M. V\"olmer,  {\it Kinetik der Phasenbildung} (Teodor
Steinkopff, Dresden und Leipzig, 1939).
\item $^{9}$ D. Turnbull, {\it J. Appl. Phys.} {\bf 21}, 1022 (1950); 
D. Turnbull and J. C. Fisher, J. Chem. Phys. {\bf 17},71 (1949).
\item $[10]$ J. Lothe and G.M.J. Pound, in A. C. 
Zettlemoyer (Ed.), Nucleation, Marcel-Dekker, New York, 1969.  
\item $^{11}$ Y. S. Djikaev, A. Tabazadeh, P. Hamill,  and H. Reiss,
J.Phys.Chem. A {\bf 106}, 10247 (2002).
\item $^{12}$ Y. S. Djikaev, A. Tabazadeh,  and H. Reiss,
J.Chem.Phys. {\bf 118}, 6572-6581 (2003).
\item $^{13}$ R. Defay, I. Prigogine, A. Bellemans, and D. H. Everett,
{\it Surface Tension and Adsorption} (John Wiley, New York, 1966).
\item $^{14}$ J. Zell and B. Mutaftshiev, Surf. Sci.
{\bf 12}, 317 (1968);
G. Grange and  B. Mutaftshiev, Surf. Sci. {\bf 47}, 723 (1975);
G. Grange, R. Landers, and B. Mutaftshiev, Surf. Sci. {\bf 54}, 445 (1976).
\item $[15]$ D.Chatain and P.Wynblatt, in {\em Dynamics of Crystal Surfaces and Interfaces}, 
Ed. P.M.Duxbury and T.J.Pence, 53-58 (Springer, NY, 2002).
\item $^{16}$ M. Elbaum, S. G. Lipson, and J. G. Dash,
J. Cryst. Growth {\bf 129}, 491 (1993).
\item $^{17}$ A.Y. Zasetsky, R. Remorov, and I.M. Svishchev, 
Chem.Phys.Let. {\bf 435} 50-53 (2007).  
\item $^{18}$ Y.G.Chushak and L.S.Bartell, J.Phys.Chem. B  {\bf 103}, 11196 (1999).
\item $^{19}$ A. Tabazadeh, Y. S. Djikaev, P. Hamill, and H. Reiss,
J. Phys. Chem. A {\bf 106}, 10238-10246 (2002).
\item $^{20}$ A. Tabazadeh, Y. S. Djikaev, and H. Reiss,
Proc. Natl. Acad. Sci. USA {\bf 99}, 15873 (2002).
\item $^{21}$R.A.Shaw, A.J.Durant, and Y.Mi, J.Phys.Chem.B {\bf 109}, 9865 (2005).
\item $^{22}$A.J.Durant and R.A.Shaw, Geophys.Res.Lett. {\bf 32}, L20814 (2005).
\item $^{23}$ J. Frenkel, {\it Kinetic Theory of Liquids} (Clarendon, Oxford, 1946).
\item $^{24}$ D. E. Sullivan and M. M. Telo da Gama, in {\it Fluid
Intefacial Phenomena}; Croxton, C. A., Ed. (John Wiley \& Sons, New
York, 1986).
\item $(25)$ MacKenzie, R.; Kulmala, M.; Laaksonen, A.; Vesala, T. 
{\it J. Geophys. Res.} {\bf 1995}, {\it 100}, 11275; 
{\it J. Geophys. Res.} {\bf 1997}, {\it 102}, 19729. 
\item $(26)$ Bertram, A. K.; Sloan, J. J. 
{\it J. Geophys. Res.} {\bf 1998}, {\it 103}, 3553; {\it J. Geophys. Res.} {\bf 1998}, 
{\it 103}, 13261. 
\item $(27)$ F.M.Kuni, Colloid J. USSR, {\bf 46}, 595-601 (1984); 
Colloid J. USSR, {\bf 46}, 791-797 (1984).
\item $(28)$ K.C.Russell, J.Chem.Phys. {\bf 50}, 1809 (1969).
\item $(29)$  A.I. Rusanov and  F.M. Kuni,  {\it Dokl. Phys. Chem.} {\bf 318}, 467-469 (1991);
{\it Colloids  Surf.} {\bf 61}, 349-351 (1991).  
\item $(30)$ Y.S.Djikaev and D.J.Donaldson, J.Geophys.Res.-Atmos. {\bf 104}, 14,283 (1999). 
\item $(31)$ Y.S.Djikaev and D.J.Donaldson, J.Geophys.Res.-Atmos. {\bf 106}, 14,447 (2001); 
J.Chem.Phys. {\bf 115}, 6822-6830 (2000). 
\item $(32)$ Kuni, F.M., A.K. Shchekin, and  A.I. Rusanov,  
Colloid J. Russ. Acad. Sci., {\bf 55}, 174-183 (1993); 
Colloid J. Russ. Acad. Sci., {\bf 55}, 202-210 (1993); 
Colloid J. Russ. Acad. Sci., {\bf 55}, 211-226 (1993).
\item $[33]$ G.Tammann, Z.Phys.Chem.Stoechiom.Verwandtschalft, {\bf 68}, 205 (1910). 
\item $[34]$ H.B.Lyon and G.A.Somorjai, J.Chem.Phys. {\bf 46}, 2539 (1967).
\item $[35]$ J.W.M.Frenken, P.M.J.Mar\'ee, and J.F. van der Veen,  Phys.Rev. B, 34, 7506, (1986).
\item $[36]$ H. Hakkinen and U. Landmann, Phys. Rev. Lett., 71, 1023, (1993) 
\item $[37]$ O. Tomagnini, F. Ercolessi, S. Iarlori, F.D. Di Tolla and E. Tosatti, 
Phys. Rev. Lett., 76, 1118, (1996) 
\item $[38]$ 
A.K. Ivanov-Schitza, G.N. Mazob, E.S. Povolotskaya, and
S.N. Savvin, Solid State Ionics {\bf 173}, 103-105  (2004).
\item $[39]$ M.A.S.M. Barrera, J.F.Sanz, L.J.\'Alvarez, and J.A.Odriozola,  
Phys.Rev.B {\bf 58}, 6057-6062 (1998).
\item $[40]$ R. Zivieri, G. Santoro, and V. Bortolani,
Phys.Rev.B {\bf 62}, 9985-9988 (2000).
\item $[41]$ F.Lindemann, Z.Phys, 11, 609, (1910)    
\item $[42]$ V. Sadtchenko and G.E.Ewing, Canad.J.Phys. {\bf 81} 333-341, (2003). 
\item $[43]$ H.Bluhm, D.F.Ogletree, C.S.Fadley, Z.Hussain, and M.Salmeron, 
J.Phys.: Condens.Matter {\bf 14}, L227-L233 (2002).
\item $[44]$ J.S.Wettlaufer, Philos.Trans.: Math.,Phys., Engin.Sci., {\bf 357}, 
3403-3425 (1999).
\item $[45]$ M.Faraday, Phil.Mag. {\bf 17} 162 (1840). 
\item $[46]$ X.Wei, P.B.Miranda, and Y.R.Shen, Phys.Rev.Lett. {\bf 86}, 1554-1557 (2001); 
X.Wei and Y.R.Shen Applied Phys. B v.74 617-620 (2002)
\item $^{47}$J.W. Cahn,  J. Chem. Phys. {\bf 66}, 3667 (1977).
\item $^{48}$S.Dietrich, in {\it Phase Transition and Critical Phenomena}, Vol.12,
p.148, eds. C. Domb and J. H. Lebowitz (Academic Press, San Diego, 1988);
\item $[49]$K.G.Libbrecht, Rep.Prog.Phys. {\bf 68}, 855-895 (2005).
\item $[50]$A.Cahoon, M.Maruyama, and J.S.Wettlaufer, Phys. Rev. Lett. {\bf 96} 255502 (2006).


\end{list}

\newpage
\subsection*{Captions} to  Figures 1 to 3 of the manuscript \\
{\sc Effect of the surface-stimulated mode on the 
kinetics of homogeneous crystal nucleation in droplets}\\
by {\bf Y. S. Djikaev}
\subsection*{}
Figure 1. A liquid droplet surrounded by vapor.
Case $a$: homogeneous crystal nucleation {\em within} the droplet (volume-based mode). Case $b$:
homogeneous crystal nucleation with one of crystal facets {\em at} the droplet surface
(surfac-stimulated mode).
\\.\\
Figure 2. Illustration to Wulff's relations (4) and (9). 
The  surface area and surface tension of the facet $i$ are denoted by $A_{i}$ and
$\sigma_{i}$, 
respectively; $h_{i}$ is the distance
from the facet $i$ to the reference point $O$ (see the text for more detail).
\\.\\
Figure 3. The nucleation mode factor $\alpha=I^{ss}/I^{vb}$  (given by eq.(36) with $w=2$) 
as a function of $R$ for
crystal nucleation in water droplets at $T=233$ K. 
The dashed and solid curves correspond to
$\gamma=2$ and $\gamma=0.5$, respectively.  
\\.\\

\newpage
\begin{figure}[htp]
\begin{center}\vspace{1cm}
\includegraphics[width=8.3cm]{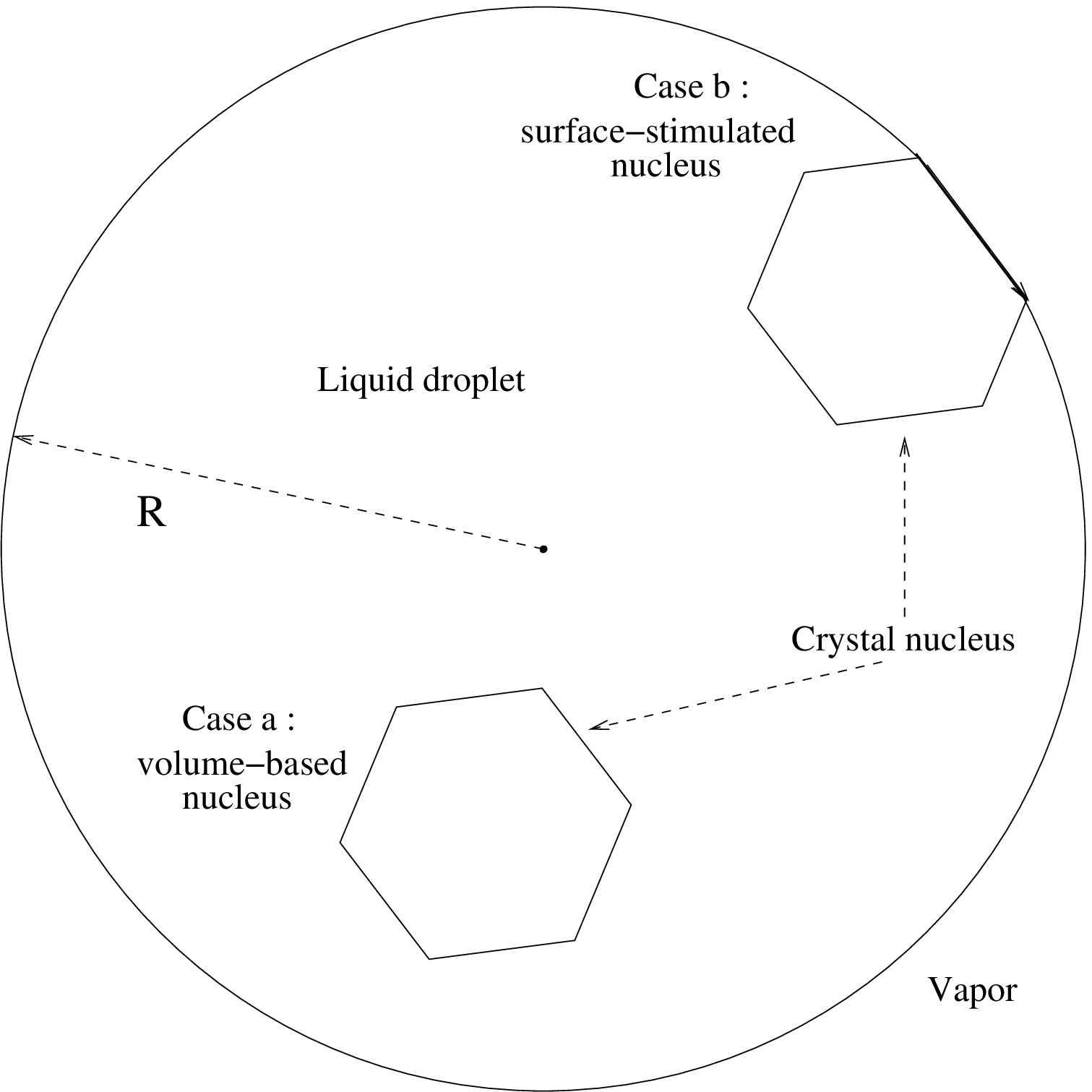}\\ [3.7cm]
\caption{\small }
\end{center}
\end{figure}

\newpage
\begin{figure}[htp]
\begin{center}\vspace{1cm}
\includegraphics[width=8.3cm]{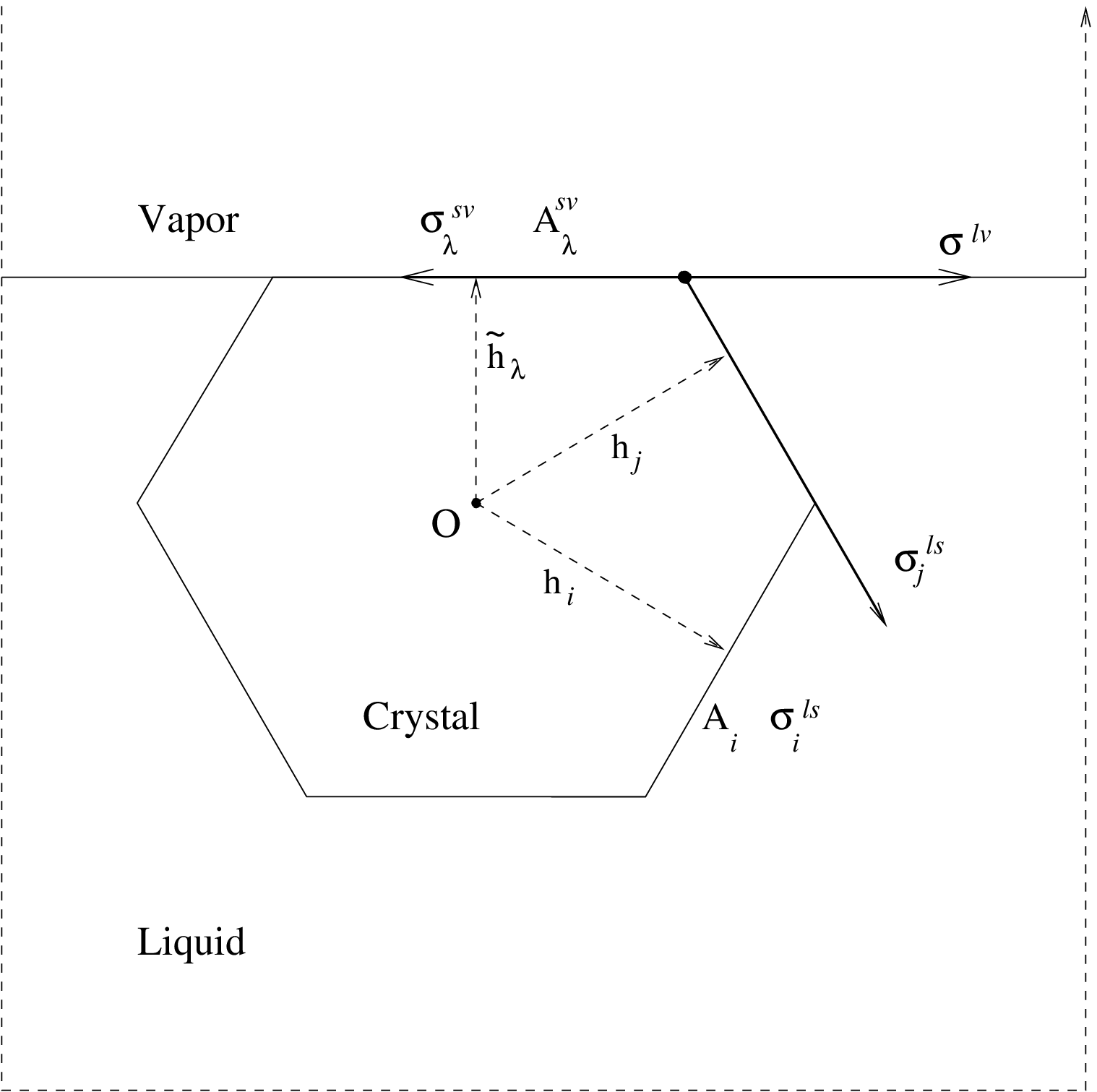}\\ [3.7cm]
\caption{\small }
\end{center}
\end{figure}

\newpage
\begin{figure}[htp]
\begin{center}\vspace{1cm}
\includegraphics[width=8.3cm]{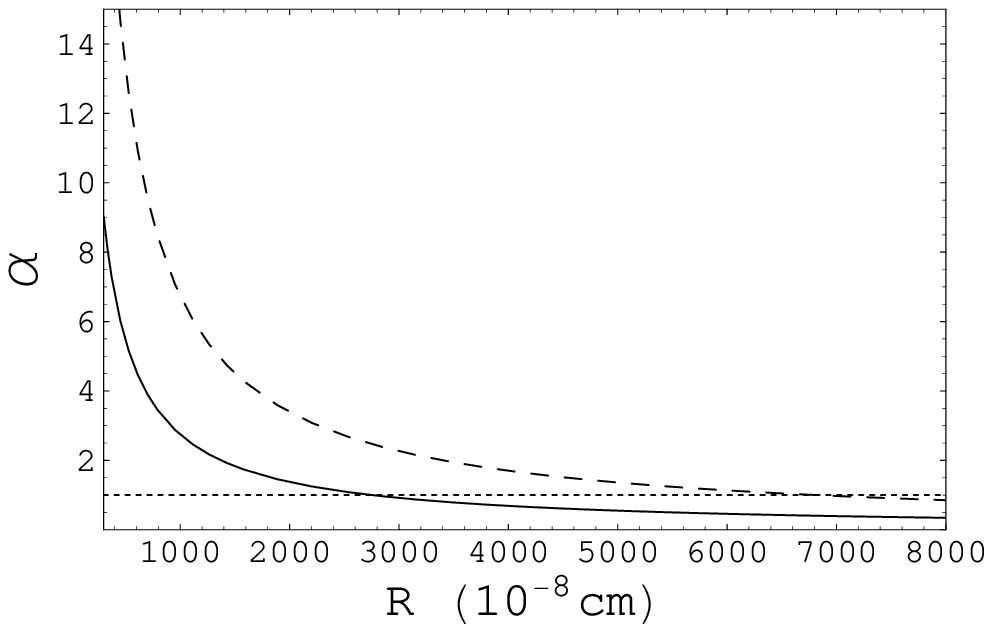}\\ [3.7cm]
\caption{\small }
\end{center}
\end{figure}

\end{document}